\documentclass[usenatbib]{mnras}
\usepackage{natbib,amssymb,amsmath,times,graphicx,url,aas_macros}
\usepackage{setspace}      
\usepackage{epstopdf}       
\usepackage{verbatim}       
\usepackage{bm}		  
\usepackage{fleqn}		  
\usepackage{Wurster_macros} 

\defcitealias{WursterBatePrice2018sd}{WBP2018}

\title[Hall effect-driven disc formation]{Hall effect-driven formation of gravitationally unstable discs in magnetized molecular cloud cores}

\author[Wurster, Bate \& Price]{James Wurster$^{1}$\thanks{j.wurster@exeter.ac.uk}, Matthew R. Bate$^{1}$\thanks{mbate@astro.ex.ac.uk}, and Daniel J. Price$^{2}$ \\
$^{1}$School of Physics and Astronomy, University of Exeter, Stocker Rd, Exeter EX4 4QL, UK \\
$^{2}$Monash Centre for Astrophysics and School of Physics and Astronomy, Monash University, Vic 3800, Australia \\
}
\date{Submitted: Revised: Accepted: }
\pagerange{\pageref{firstpage}--\pageref{lastpage}} \pubyear{2018}

\begin{document}
\label{firstpage}
\bibliographystyle{mnras}
\maketitle

\begin{abstract}
We demonstrate the formation of gravitationally unstable discs in magnetized molecular cloud cores with initial mass-to-flux ratios of 5 times the critical value, effectively solving the magnetic braking catastrophe.  We model the gravitational collapse through to the formation of the stellar core, using Ohmic resistivity, ambipolar diffusion and the Hall effect and using the canonical cosmic ray ionization rate of $\zeta_\text{cr} = 10^{-17}$ s$^{-1}$.  When the magnetic field and rotation axis are initially aligned, a $\lesssim1$~au disc forms after the first core phase, whereas when they are anti-aligned, a gravitationally-unstable 25~au disc forms during the first core phase.  The aligned model launches a 3~km~s$^{-1}$ first core outflow, while the anti-aligned model launches only a weak $\lesssim 0.3$~km~s$^{-1}$ first core outflow.  Qualitatively, we find that models with $\zeta_\text{cr} = 10^{-17}$ s$^{-1}$ are similar to purely hydrodynamical models if the rotation axis and magnetic field are initially anti-aligned, whereas they are qualitatively similar to ideal magnetohydrodynamical models if initially aligned.
\end{abstract}

\section{Introduction}
\label{intro}

Molecular clouds are magnetized \citep[for a review, see][]{HeilesCrutcher2005} but with low ionization fractions \citep{MestelSpitzer1956,NakanoUmebayashi1986,UmebayashiNakano1990}.  The canonical cosmic ray ionization rate in molecular clouds is $\zeta_\text{cr} \approx 10^{-17}$ s$^{-1} \exp\left(-\Sigma/\Sigma_\text{cr}\right)$ \citep{SpitzerTomasko1968,UmebayashiNakano1981}, where $\Sigma$ is the surface density of the gas, and $\Sigma_\text{cr}$ is the characteristic attenuation depth of cosmic rays.  The dense regions ultimately collapse to form protostars \citep{Shu1977}, and observations have suggested the presence of large gas discs and outflows around these young (Class 0) objects \citep[e.g.][]{Dunham+2011,Lindberg+2014,Tobin+2015,Gerin+2017}.

Despite the low ionization fractions, many recent simulations of magnetized star formation used ideal magnetohydrodynamics \citep[MHD; e.g.][]{PriceBate2007,HennebelleFromang2008,DuffinPudritz2009,HennebelleCiardi2009,Commercon+2010,Seifried+2011,BateTriccoPrice2014},  which assumes that the gas is sufficiently ionized such that the magnetic field is `frozen' into the gas.  The simulations that included realistic magnetic field strengths (mass-to-flux ratios of 3--5 times critical) produced collimated outflows but no protostellar discs; the lack of discs is known as the magnetic braking catastrophe \citep[e.g.][]{AllenLiShu2003,PriceBate2007,MellonLi2008,HennebelleCiardi2009}.  The simulations that included weak magnetic fields ($\gtrsim$10 times critical mass-to-flux ratio) produced weak outflows and large discs during the first hydrostatic core phase.  If large discs rotated rapidly enough, then they could become dynamically unstable to a bar-mode instability, leading to the formation of trailing spiral arms, as seen in purely hydrodynamical simulations \citep[e.g.][]{Bate1998,SaigoTomisaka2006,SaigoTomisakaMatsumoto2008,MachidaInutsukaMatsumoto2010,Bate2010,Bate2011}.

In attempts to form discs during the star forming process, recent three-dimensional simulations have accounted for the low ionization fractions by including a self-consistent treatment of non-ideal MHD \citep[e.g.][]{MachidaMatsumoto2011,Tomida+2013,TomidaOkuzumiMachida2015,Tsukamoto+2015oa,Tsukamoto+2015hall,WursterPriceBate2016,Tsukamoto+2017,Vaytet+2018,WursterBatePrice2018sd}.  Rotationally supported discs have been found in simulations that include Ohmic resisitivity and/or ambipolar diffusion \citep[e.g.][]{TomidaOkuzumiMachida2015,Tsukamoto+2015oa,Vaytet+2018}, and 15-30~au discs were recovered when the Hall effect was included \citep[e.g.][]{Tsukamoto+2015hall,WursterPriceBate2016,Tsukamoto+2017} so long as the magnetic field was anti-aligned with the rotation axis, since this geometry promotes disc formation \citep{BraidingWardle2012accretion}.

\newpage
In this paper, which follows from the work presented in \citet{WursterBatePrice2018sd} (hereafter \citetalias{WursterBatePrice2018sd}), we model the gravitational collapse of a magnetised molecular cloud core using Ohmic resistivity, ambipolar diffusion and the Hall effect and the canonical cosmic ray ionization rate of \zetaeq{-17}.  This is the first study to model the collapse to the stellar core phase \citep{Larson1969} that includes the three main non-ideal effects, uses the canonical cosmic ray ionisation rate of \zetaeq{-17}, and anti-aligns the initial magnetic field and rotation vectors.  Previous studies have anti-aligned the vectors but stopped the evolution after the first core phase \citep{Tsukamoto+2015hall,Tsukamoto+2017}; aligned the vectors and evolved to the stellar core phase 
\citep{Tsukamoto+2015hall}; 
studied both alignments using a higher cosmic ray ionization rate \citepalias{WursterBatePrice2018sd}; excluded the Hall effect  \citep{TomidaOkuzumiMachida2015,Tsukamoto+2015oa,Vaytet+2018}; or followed the long term evolution by forming sink particles \citep{WursterPriceBate2016}.

This paper focuses on disc formation. We refer the reader to \citetalias{WursterBatePrice2018sd} 
for discussion of the stellar cores.  In Section~\ref{sec:methods}, we summarise our methods and in Section~\ref{sec:ic} we present our initial conditions.  Our results are presented in Section~\ref{sec:results} and we conclude in Section~\ref{sec:conc}.

\section{Methods}
\label{sec:methods}

Our method is almost identical to that used by \citetalias{WursterBatePrice2018sd}. We solve the equations of self-gravitating, radiation non-ideal magnetohydrodynamics using the three-dimensional smoothed particle hydrodynamics (SPH) code \textsc{sphNG} that originated from  \citet{Benz1990}, but has since been substantially modified to include a consistent treatment of variable smoothing lengths \citep{PriceMonaghan2007}, individual timestepping \citep{BateBonnellPrice1995}, radiation as flux limited diffusion \citep{WhitehouseBateMonaghan2005,WhitehouseBate2006}, magnetic fields \citep[for a review, see][]{Price2012}, and non-ideal MHD \citep{WursterPriceAyliffe2014,WursterPriceBate2016} using the single-fluid approximation.  

For stability of the magnetic field, we use the \citet{BorveOmangTrulsen2001} source-term approach, maintain a divergence-free magnetic field using constrained hyperbolic/parabolic divergence cleaning \citep{TriccoPrice2012,TriccoPriceBate2016}, and use the artificial resistivity from \citet{Phantom2017}; note that \citetalias{WursterBatePrice2018sd} used the artificial resistivity from \citet{TriccoPrice2013}.

The non-ideal MHD coefficients \citep[for review, see][]{Wardle2007} are calculated using Version 1.2.1 of the \textsc{Nicil} library \citep{Wurster2016} using its default values.  We include Ohmic resistivity, ambipolar diffusion and the Hall effect.  At low temperatures ($T \lesssim 600$~K), cosmic rays are the dominant ionisation source and can ionise low mass ions ($m \sim 2.31$m$_\text{p}$, where m$_\text{p}$ is the proton mass), high-mass ions ($m \sim 24.3$m$_\text{p}$), and dust grains; the mutual interaction between the species can lead to further ionisation.  The dust grains are comprised of a single species with radius $a_\text{g}=0.1\mu$m, bulk density $\rho_\text{bulk}=3$~\gpercc, and a dust-to-gas fraction of 0.01 \citep{Pollack+1994}.  The dust is evolved as three populations, which are negatively, positively, and neutrally charged, respectively.

\section{Initial conditions}
\label{sec:ic}

Our initial conditions are identical to those in \citet{BateTriccoPrice2014} and \citetalias{WursterBatePrice2018sd}.  We place a 1~M$_{\odot}$ cold dense sphere of uniform density into warm background at a density ratio of 30:1, with the two phases in pressure equilibrium.  The core has an initial radius of $R = 4\times 10^{16}$~cm, an initial sound speed of  $c_\text{s} = \sqrt{p/\rho} =  2.2\times 10^{4}$~cm~s$^{-1}$, and solid body rotation about the $z$-axis of $\bm{\Omega}_0 = \Omega_0\hat{\bm{z}}$, where $\Omega_0 = 1.77 \times 10^{-13}$~rad s$^{-1}$.   

The entire domain is threaded with a magnetic field in the $z$-direction with a strength of $B_0 = 1.63 \times 10^{-4}$~G, equivalent to 5 times the critical mass-to-flux ratio.  For the models that include non-ideal MHD, we consider both aligned (i.e. \BBpz) and anti-aligned (i.e. \BBnz) cases. 
 
The calculations use $3 \times 10^{6}$ equal-mass SPH particles in the core and $1.46 \times 10^{6}$ particles in the surrounding medium. 

\section{Results}
\label{sec:results}

Our suite of simulations consists of a purely hydrodynamical model (named HD), an ideal MHD model (named iMHD), and four non-ideal MHD models named \zetam{Z}{D}, where $Z$ represents the constant cosmic ray ionization rate such that \zetaeq{-Z}, and $D$ represents the direction of the magnetic field with respect to the rotation axis such that $D = + \ (-)$ represents \Bpz \ (\Bnz).  Our focus is on \zetam{17}{\pm}, and the other models are included for reference and to emphasise the importance of a proper treatment of non-ideal MHD.

\subsection{Column density evolution}

\begin{figure*}
\centering
\includegraphics[width=0.8\textwidth]{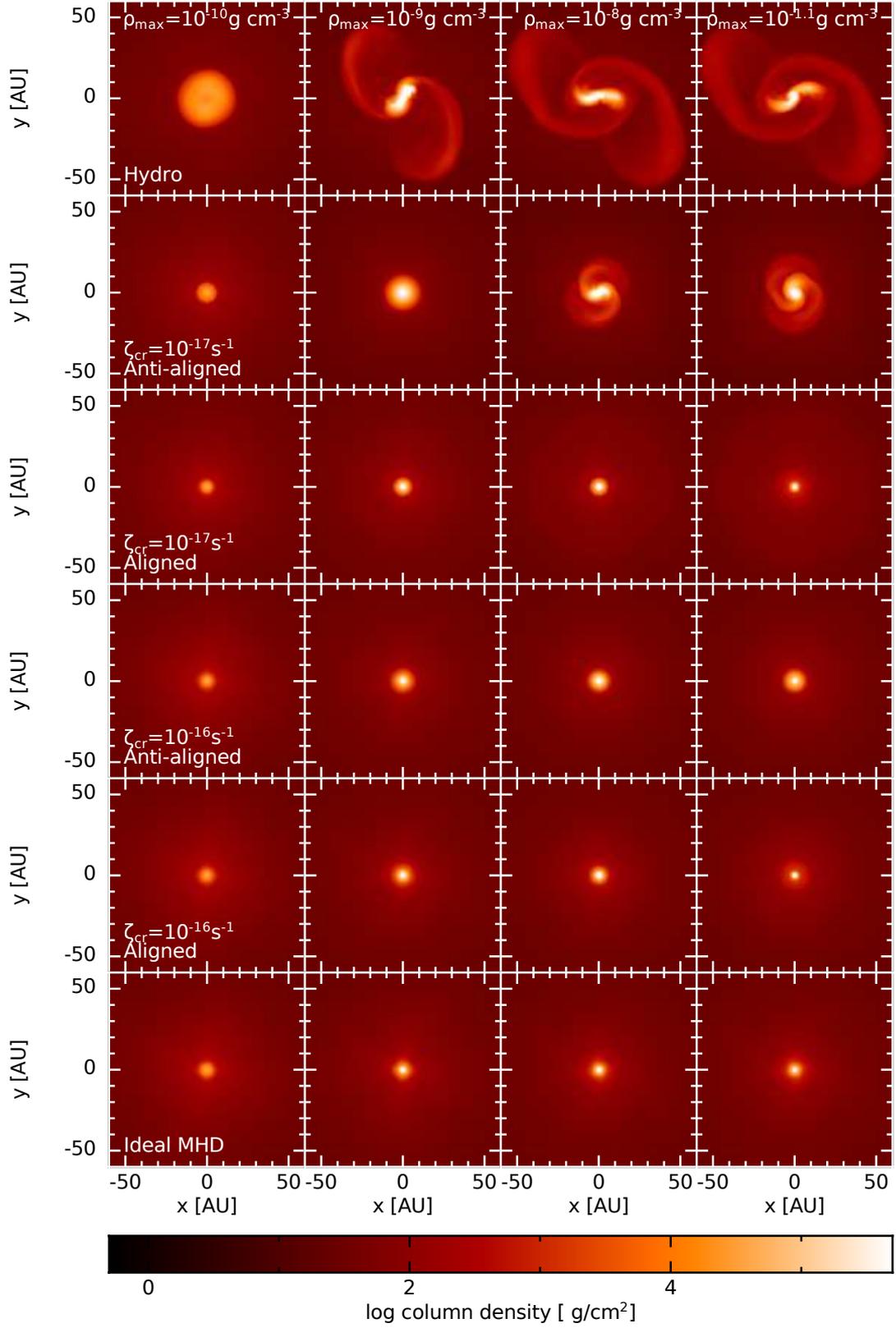} 
\caption{Formation of gravitationally unstable discs in the presence of magnetic fields, showing the face-on gas column density at selected maximum densities (a proxy for time).  The hydrodynamic model (top row) forms a \sm60~au disc that becomes bar-unstable and forms spiral arms, while the ideal MHD model (bottom row) forms no disc.  The Hall effect in model \zetamp{17} prevents disc formation (third row), whereas the Hall effect increases the angular momentum contained in the disc in model \zetamn{17} (second row) to allow a gravitationally unstable \sm25~au disc to form.}
\label{fig:rho}
\end{figure*}

Fig.~\ref{fig:rho} shows the face-on gas column density during the first and stellar core phases.  As expected, a gravitationally unstable disc forms ($r \sim 60$~au) in the HD calculation that undergoes a gravitational bar-mode instability \citep{Bate1998,Durisen+2007} early in the first core phase, while no disc forms in iMHD in agreement with \citet{BateTriccoPrice2014}.  In agreement with \citet{Tsukamoto+2015hall}, \zetamn{17} develops a large \sm25~au disc during the first core phase that becomes gravitationally unstable and forms spiral arms.  By reversing the initial direction of the magnetic field, \zetamp{17} forms no disc during the first core phase.  As the collapse proceeds from the first hydrostatic core to the stellar core, the bars in HD and \zetamn{17} collapse to form a spherical core.

Our study adopts different initial rotations, different initial magnetic field strengths and different microphysics governing the non-ideal MHD processes compared to \citet{Tsukamoto+2015hall}, suggesting that the bimodality of disc formation is robust to initial conditions.  That is, for models with \zetaeq{-17}, the evolution is similar to HD if the initial magnetic field and rotation vectors are anti-aligned, and similar to iMHD if the vectors are aligned. 

\subsection{Angular momentum and magnetic braking}
\begin{figure}
\centering
\includegraphics[width=\columnwidth]{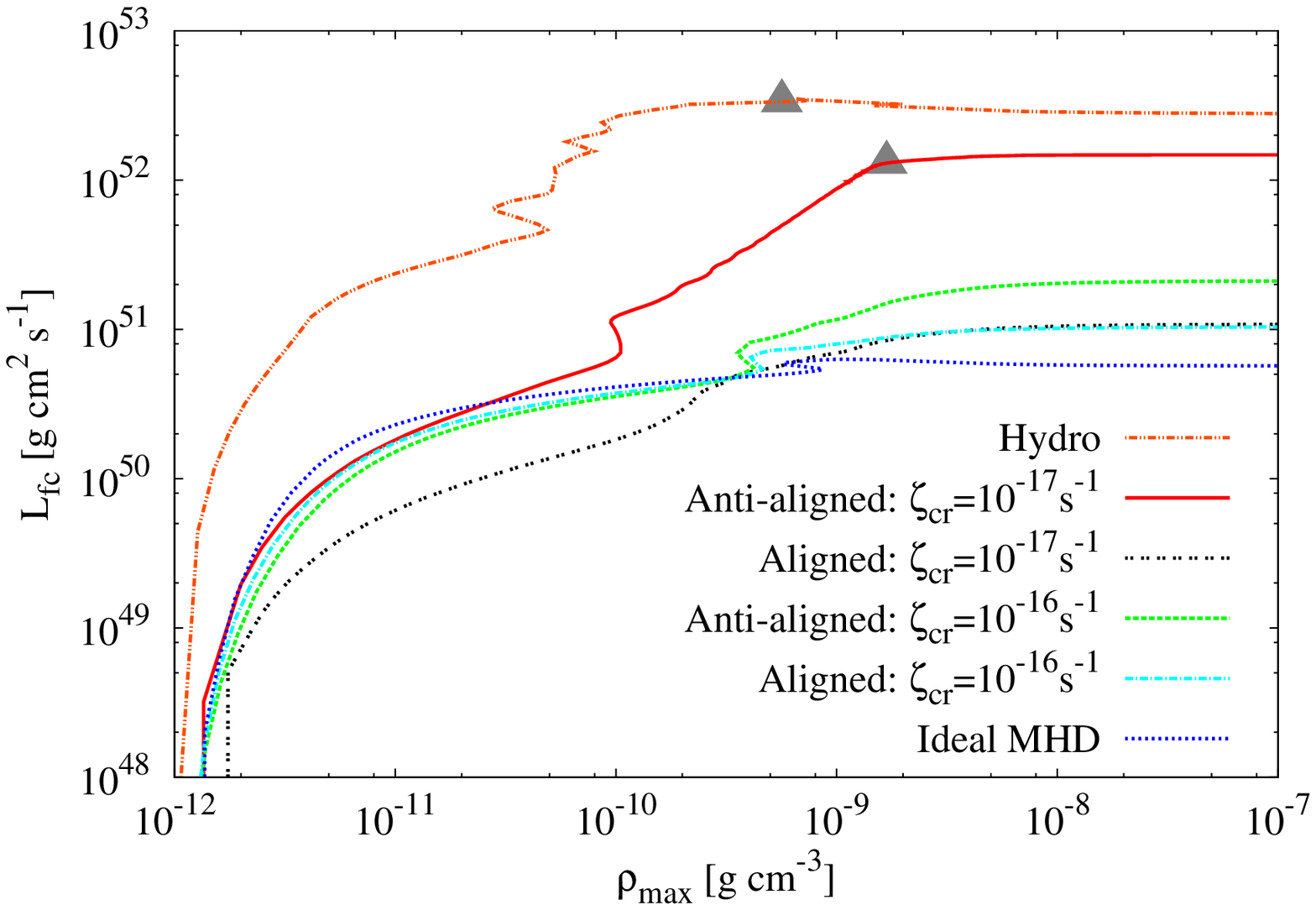}  \vspace{-0.25cm}
\caption{Evolution of the angular momentum in the first hydrostatic core (defined as the gas with \rhoge{-12}).  The triangles represent when the discs becomes gravitationally unstable.   The angular momentum in the first core is larger for models with lower ionization rates, and with initially anti-aligned magnetic field and rotation vectors.}
\label{fig:fhc:LVrho}
\end{figure}

In the purely hydrodynamic calculation, conservation of angular momentum during the initial collapse to form the first hydrostatic core results in the formation of a gravitationally unstable disc of radius \sm60~au, as shown in the top row of Fig.~\ref{fig:rho}.  Fig.~\ref{fig:fhc:LVrho} shows the evolution of the angular momentum in the first core, $L_\text{fc}$, where the first core is defined as all the gas with \rhoge{-12}.   Magnetic fields are efficient at transporting angular momentum outwards, thus the first core in iMHD has \sm50 times less angular momentum than HD.  As a result, a rotationally supported disc does not form in iMHD.

Angular momentum directly affects the time evolution of the collapse. The length of time spent in the first core phase increases as the angular momentum of the first core increases.  The exception to this trend is \zetamn{17}, which has an even longer first core phase (\smq630~yr) than the HD model (\smq590~yr) despite having slightly less angular momentum.  This is because in \zetamn{17} the magnetic field supports the cloud against gravity and delays the collapse, extending the lifetime of the first hydrostatic core phase.

In all simulations, total angular momentum is conserved within 1 per cent during the entire gravitational collapse through to stellar densities.  The initial angular momentum in our simulations is $2.26\times10^{53}$~g~cm$^2$~s$^{-1}$, and, in \zetamn{17}, \sm6.5 per cent of this is contained in first hydrostatic core after its formation.

\subsubsection{Ion and bulk velocities}
\begin{figure}
\centering
\includegraphics[width=\columnwidth]{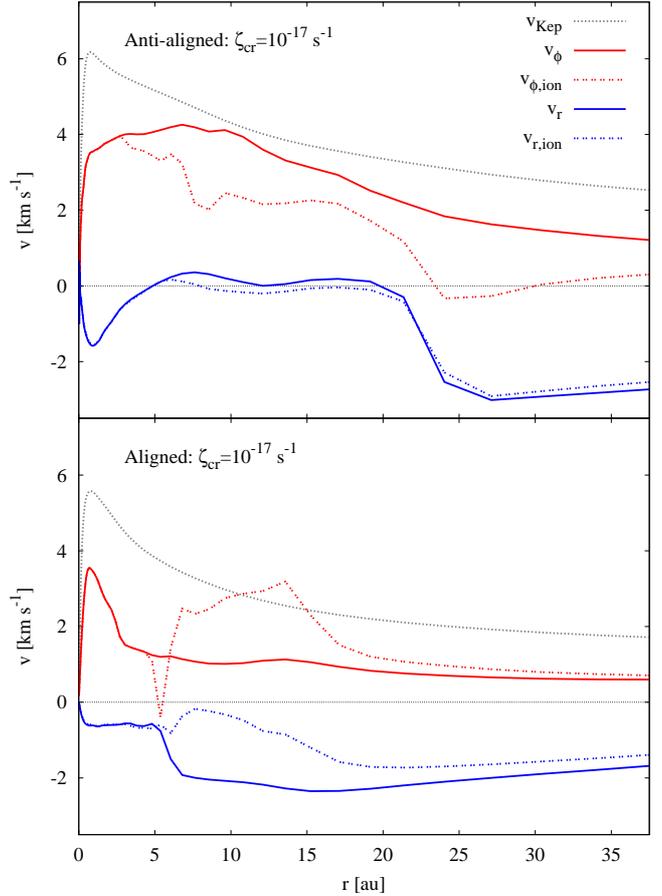} 
\caption{Azimuthally averaged single-fluid and ion velocities within $20^\circ$ of the midplane at \rhoxapprox{-7} for models \zetamn{17} (top) and \zetamp{17} (bottom).  The gas is rotating at sub-Keplerian velocities.  The ions are rotating slower than the bulk rotational flow in \zetamn{17}, decreasing the magnetic braking and promoting disc formation.}
\label{fig:vel}
\end{figure}

Fig.~\ref{fig:vel} shows the azimuthally averaged radial and azimuthal velocities, $v_\text{r}$ and $v_\phi$, respectively, of both the ions and the single-fluid motion within $20^\circ$ of the midplane at \rhox \ \appx \ $10^{-7}$ for models \zetampn{17}.  The ion velocity is given by
\begin{equation}
\bm{v}_\text{ion} = \bm{v} + \frac{\rho_\text{n}}{\rho}\left(\eta_\text{A}\frac{\bm{J}\times \bm{B}}{|B|^2} - \eta_\text{H}\frac{\bm{J}}{|B|}\right),
\end{equation}
where $\bm{v}$ is the single-fluid velocity calculated in the simulations, $\rho_\text{n}$ and $\rho$ are the neutral and total mass densities, respectively, $\bm{J}$ is the current density, and $\eta_\text{H}$ and $\eta_\text{A}$ are the coefficients for the Hall effect and ambipolar diffusion, respectively.

As the ionization rate is decreased, the coupling between the matter and magnetic fields decreases.  In \zetamn{17}, the ions rotate slower than the bulk rotational flow.  This results in decreased magnetic braking and a torque that spins up the material in the same direction as the initial flow \citep{KrasnopolskyLiShang2011}, such that the angular momentum is approximately half of that in model HD.  This promotes disc formation.  

In \zetamp{17}, the ions rotate faster than the bulk rotational flow, dragging the magnetic field more rapidly around the disc.  This creates a stronger toroidal magnetic field, which enhances magnetic braking and prevents the formation of a Keplerian disc.  By reversing the direction of the initial magnetic field such that the magnetic field and rotation are initially aligned (i.e. \zetamn{17} $\rightarrow$ \zetamp{17}), the angular momentum in the first core decreases by a factor of \sm12.  

Similar trends hold for \zetampn{16} (not shown), although the difference between the ion and bulk velocities is smaller than in \zetampn{17}.  In these models, the ionisation rate is high enough to modify the rotational profile, but not enough to reduce magnetic braking enough for a disc to form during this phase.  Both \zetamp{17} and \zetamn{16} have similar angular momenta in the first core, indicating that both the cosmic ray ionisation rate and the initial magnetic field orientation are critical in determining the angular momentum content of the first core and hence disc formation.

\subsubsection{Degree of centrifugal support of the discs}
\begin{figure}
\centering
\includegraphics[width=\columnwidth]{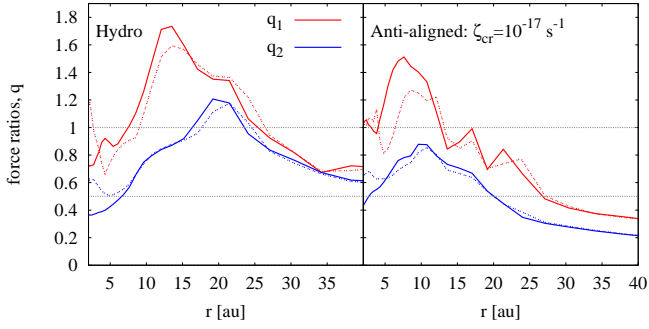}
\caption{Ratio of centrifugal and pressure forces to the gravitational force ($q_1$ and $q_2$ as defined in Eqns.~\ref{eq:q1} and \ref{eq:q2}) for models that form discs during the first hydrostatic core phase.  $q_1$ gives the ratio of the combination of the centrifugal and pressure forces to the radial gravitational force, while $q_2$ gives only the ratio of the centrifugal force to the radial gravitational force.  The forces are computed for the gas within $20^\circ$ of the midplane at \rhoxapprox{-7} (solid) and \rhoxtwoapprox{8}{-2} (dashed).  The horizontal lines are reference lines.  At both densities, the disc in \zetamn{17} is rotationally supported, with the primary contribution from the centrifugal force.} 
\label{fig:q:long}
\end{figure}

\begin{figure}
\centering
\includegraphics[width=\columnwidth]{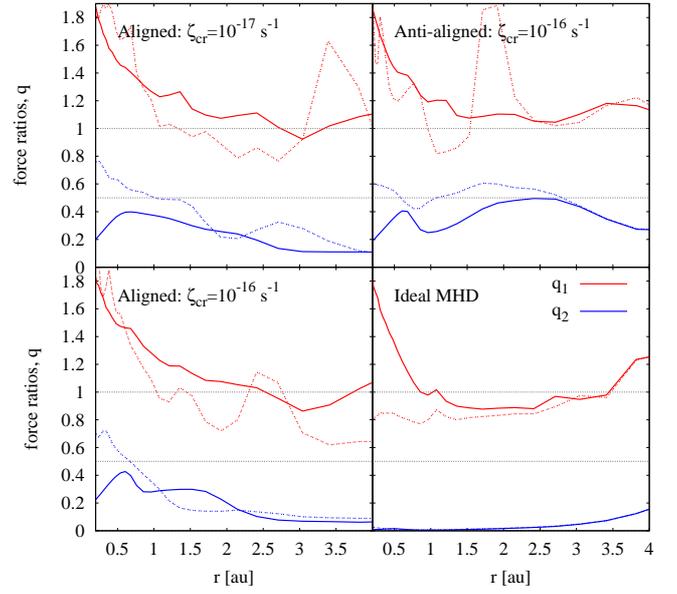} 
\caption{As in Fig.~\ref{fig:q:long}, but for models that do not form discs during the first hydrostatic core phase.  At \rhoxtwoapprox{8}{-2}, there are rotationally supported discs in the non-ideal MHD models; a rotationally supported disc does not form in iMHD.} 
\label{fig:q:short}
\end{figure}

To determine if the gas is rotationally supported, we consider the ratio of centrifugal and pressure forces to the gravitational force, namely
\begin{equation}
\label{eq:q1}
q_1 = \left| \frac{\frac{v_\phi^2}{r} + \frac{1}{\rho}\frac{\text{d}P}{\text{d}r}}{\frac{GM(r)}{r^2}}\right|,
\end{equation}
and the ratio of centrifugal force to the radial gravitational force,
\begin{equation}
\label{eq:q2}
q_2 = \left| \frac{\frac{v_\phi^2}{r} }{\frac{GM(r)}{r^2}}\right|,
\end{equation}
where $P$ is gas pressure, $M(r)$ is the mass enclosed at radius $r$, and $G$ is Newton's gravitational force constant  \citep[e.g.][]{Tsukamoto+2015oa,Tsukamoto+2015hall}.  The ratios $q_1$ and $q_2$ are shown Fig.~\ref{fig:q:long} for HD and \zetamn{17} and Fig.~\ref{fig:q:short} for the remaining models; note that each figure has a different horizontal range.  

By the end of the first core phase (at \rhoxapprox{-7}), HD and \zetamn{17} have $q_1 > 1$ for $r \lesssim 25$ and 15~au, respectively, hence discs exist that are supported against gravity, and are in close to Keplerian rotation (Fig.~\ref{fig:vel}).  Since $q_2> 0.5$, the disc is primarily supported by the centrifugal force (Fig.~\ref{fig:q:long}).   These values are smaller than previously presented since the azimuthal averaging removes information about the extended spiral arms, which were considered in our previous estimate of the disc size.  

At this \rhox, none of the remaining models have rotationally supported discs.  When we evolve the models through to the stellar core phase, then small, rotationally supported discs form in the remaining non-ideal MHD models.  These discs are $r \approx 1-3$~au in radius (depending on ionization rate and magnetic field orientation), and are also primarily supported by the centrifugal force.  Model iMHD has $q_2 \approx 0$, thus there is essentially no rotational support (Fig.~\ref{fig:q:short}).  Our results for \zetampn{17} are in agreement with \citet{Tsukamoto+2015hall}. 


The disc in \zetamn{17} is \sm10~au larger than we found in \citet{WursterPriceBate2016}.  This is a result of our previous study using sink particles (which remove gas pressure of the central region) and a barotropic equation of state.  We performed a set of additional simulations (not shown), and verified that models that use smaller sink particles form larger and more dense discs (when using the barotropic equation of state), and that models that use radiation hydrodynamics form larger discs than those using the barotropic equation of state.

\subsection{Magnetic field evolution}
\begin{figure}
\centering
\includegraphics[width=\columnwidth]{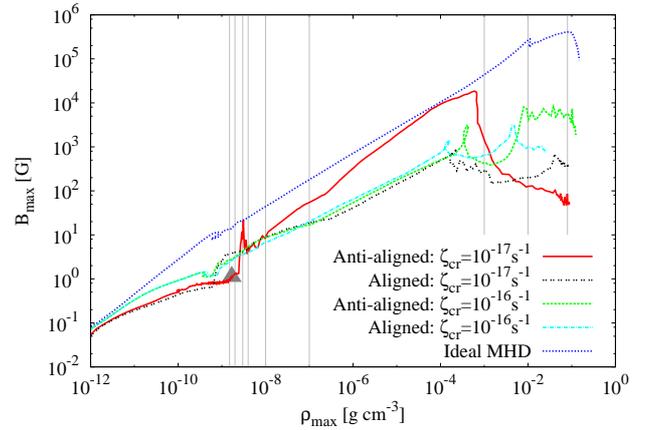} \vspace{-0.25cm}
\caption{Evolution of the maximum magnetic field strength as a function of maximum density.  The horizontal axis begins near the start of the first core phase, when the magnetic field evolution diverges amongst the models.  The grey reference lines correspond to the maximum densities shown in Figs.~\ref{fig:bar:rho} and \ref{fig:bar:B}.  The triangle represents when the disc becomes gravitationally unstable.  The maximum magnetic field strength in \zetamn{17} increases by an order of magnitude at \rhoxtwoapprox{2}{-9}, near which time the disc becomes bar-unstable.}
\label{fig:BVrho}
\end{figure}
\begin{figure*}
\centering
\includegraphics[width=0.8\textwidth]{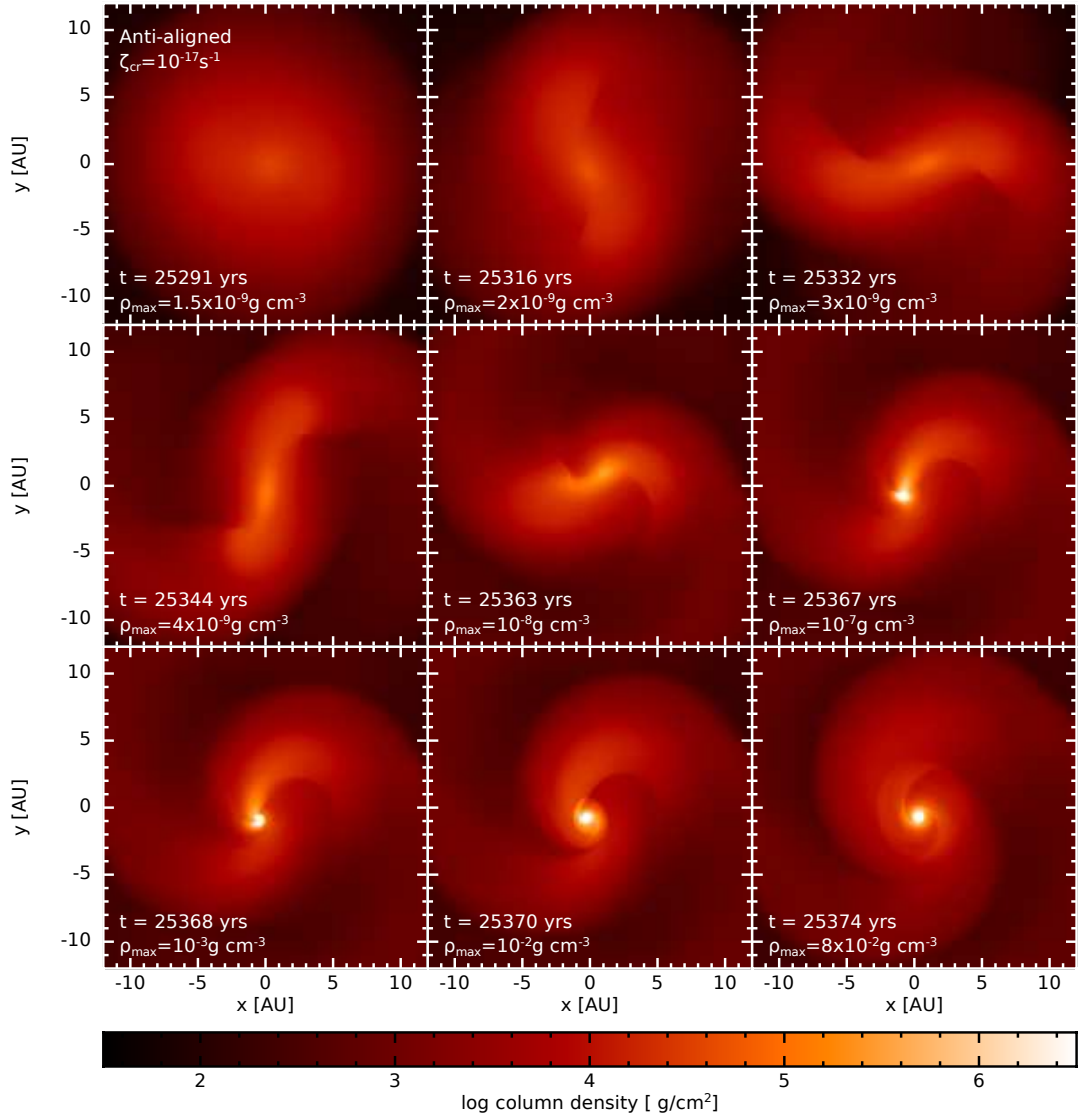} 
\caption{Evolution of the gas column density for \zetamn{17}.  The frames are chosen to highlight the bar formation and collapse.  The bar forms half-way through the first core phase, and begins to collapse by the end of this phase, ultimately forming a spherical stellar core.}
\label{fig:bar:rho}
\end{figure*}
\begin{figure*}
\centering
\includegraphics[width=0.8\textwidth]{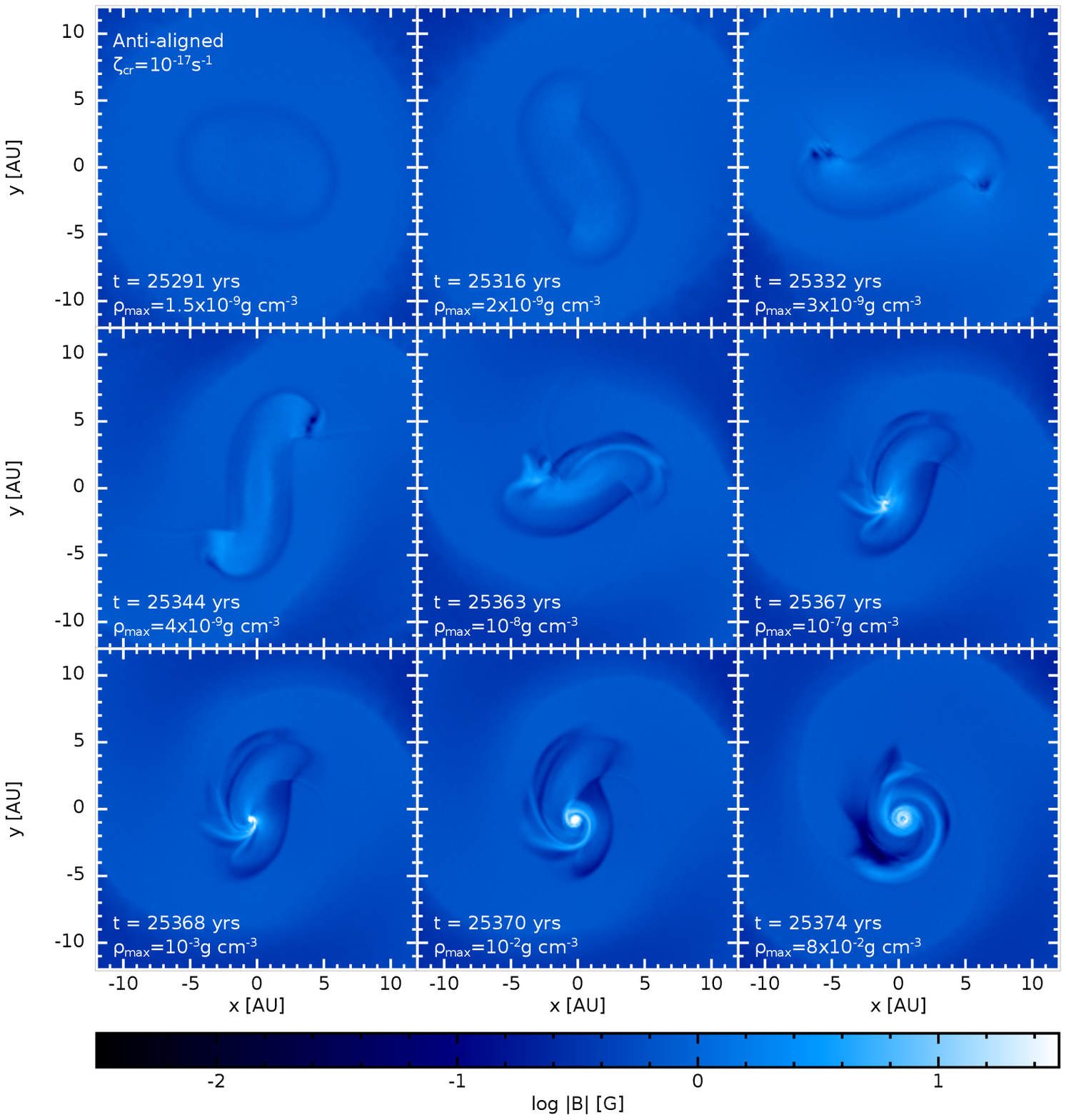} 
\caption{Evolution of the magnetic field strength in the midplane for \zetamn{17}.  The frames are chosen to highlight the bar formation and collapse.  The maximum magnetic field and maximum density are coincident only up until bar formation.}
\label{fig:bar:B}
\end{figure*}

Fig.~\ref{fig:BVrho} shows the evolution of the maximum magnetic field strength with respect to maximum density.  After the formation of the first core, the magnetic field is diffused out of the core in the non-ideal MHD models, such that at \rhoxapprox{-9}, the maximum magnetic field strength in the non-ideal models is approximately an order of magnitude lower than in the ideal MHD model.
During the second collapse phase, the maximum magnetic field strengths grow as $B_\text{max} \propto \rho_\text{max}^{0.6}$ \citep[in agreement with, e.g.,][]{BateTriccoPrice2014,Tsukamoto+2015oa,Masson+2016,WursterBatePrice2018sd}.  

The lack of azimuthal symmetry in \zetamn{17} necessarily produces a more complex magnetic field structure.  Figs.~\ref{fig:bar:rho} and \ref{fig:bar:B} show the evolution of the gas column density and magnetic field strength, respectively, of \zetamn{17}. We show slices in the x-y plane (that is, perpendicular to the rotation axis) with times chosen to highlight the formation and collapse of the bar.  Asymmetries form during the first core phase, and a bar forms by \rhoxtwoapprox{2}{-9}; at this time, the magnetic field becomes concentrated at the ends of the bar, accounting for the sudden increase in $B_\text{max}$ shown in Fig.~\ref{fig:BVrho}.  At the end of the bar, the diffusion timescale, $t_\eta \sim r^2/\eta$, is \sm$7\times10^3$~yr assuming $r \sim 1$~au and $\eta\sim10^{15}$~cm$^2$~s$^{-1}$. The latter value is representative of the physical resistivity values in the outer region of the bar.  Nearer the centre of the core, the diffusion timescale is even longer.  This diffusion timescale is longer than the evolutionary timescale of the bar (\appx80~yr; Figs.~\ref{fig:bar:rho} and \ref{fig:bar:B}), implying that the concentration of the magnetic field at the ends of the bar cannot be rapidly diffused away.  

As the bar evolves, gravitational torques funnel the gas along the bar (e.g. third column in the middle row of Figs.~\ref{fig:bar:rho} and \ref{fig:bar:B}), forming a compact core surrounded by a large disc; this is ultimately where the stellar core forms.  As expected, the magnetic field becomes concentrated in this compact core, but is highly structured (Fig.~\ref{fig:bar:B}).

\subsection{Outflows}

\begin{figure*}
\centering
\includegraphics[width=0.8\textwidth]{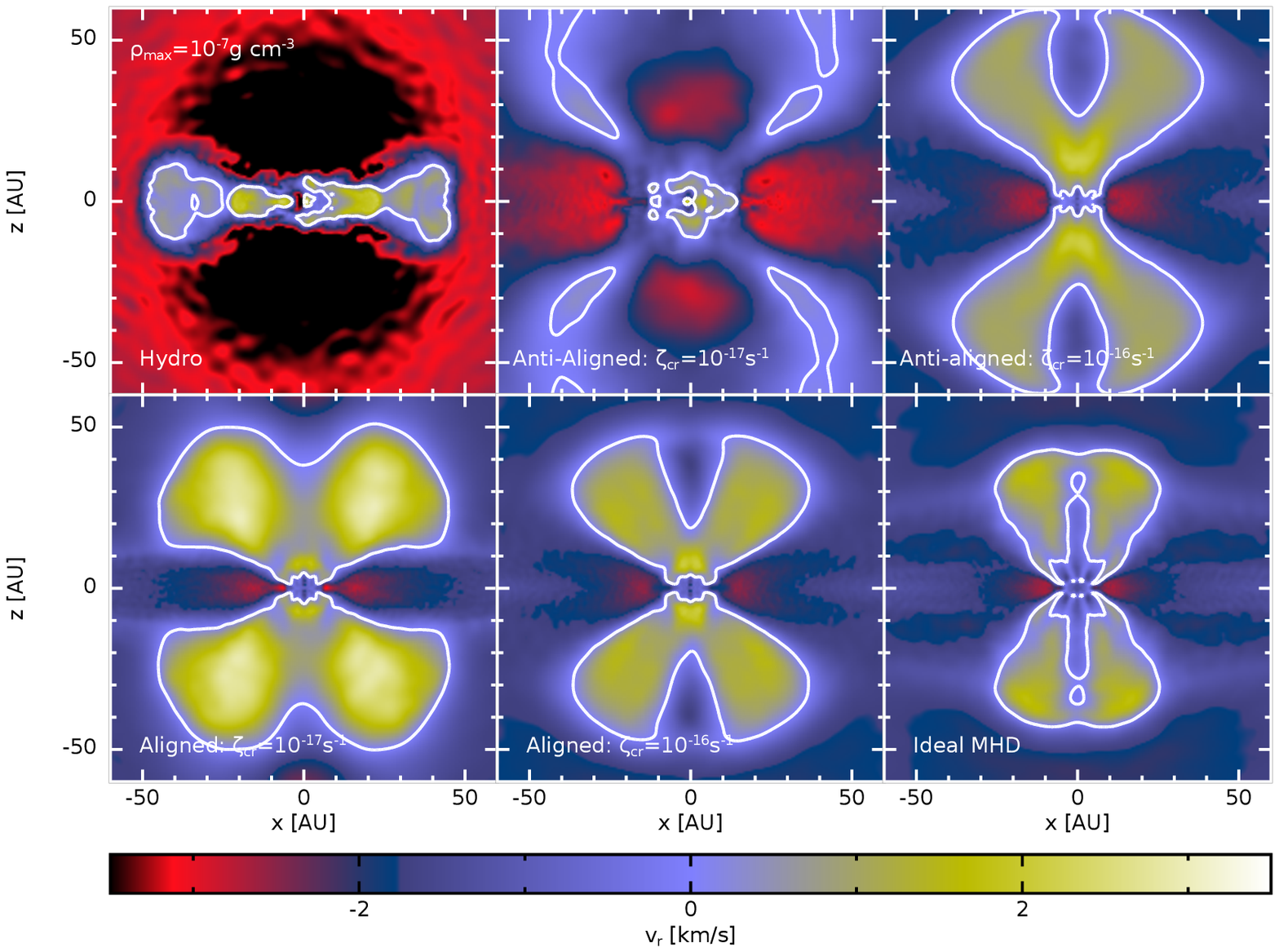} 
\caption{Radial velocity in a slice through the first core for each of the calculations.  The white contour is $v_\text{r} = 0$.  There are no first core outflows in the purely hydrodynamics model, and there are narrow, slow outflows embedded in the centre of the diagonal infall of \zetamn{17}.}
\label{fig:vr}
\end{figure*}

As shown in \citetalias{WursterBatePrice2018sd}, decreasing the cosmic ray ionization rate in models with \Bnz \ decreases the speed of the first core outflows and broadens them.  In agreement with this trend, \zetamn{17} shows a slow first core outflow; see Fig.~\ref{fig:vr}, which shows the radial velocity in a slice through the first core at \rhoxapprox{-7}.  Although a similar `X'-shaped pattern is visible for all the non-ideal MHD models, the pattern in \zetamn{17} has only a narrow band of gas that is slowly outflowing ($v_\text{r} \lesssim 0.3$~\kms), while the surrounding material is falling at a slightly faster rate; the gas is accreting faster along both the equatorial and polar directions than along the diagonals.  
Given the high angular momentum in the disc of  \zetamn{17}, outflows are not required to carry angular momentum away.

By contrast, as the ionization rate is decreased for the models with \Bpz, the speed of the first core outflows increases and broadens.  In all cases, the lower ionization rate reduces the magnetic field strength which accounts for the broadening of the outflows.  Thus, at any given \zetacr, the difference in outflow speed is necessarily a result of the Hall effect\footnote{For models with \zetage{-15}, the initial direction of the magnetic field does not significantly affect the structure or velocity of the first core outflow.}.  In the models with \Bpz, the Hall effect spins down the gas above and below the first core, which reduces the toroidal component of the magnetic field.  As previously shown (e.g. \citealp{BateTriccoPrice2014}, \citetalias{WursterBatePrice2018sd}), lower ratios of toroidal-to-poloidal magnetic field strengths result in faster outflows.  Shortly after the formation of the first core, the trend of decreasing ratios of toroidal-to-poloidal magnetic field strengths corresponds to increasing outflow speeds.

\citet{Vaytet+2018} modelled the collapse through the first and stellar core phases using \zetaeq{-17}, however, their model included only ambipolar diffusion and Ohmic resistivity.  They found no first core outflows, akin to our \zetamn{17} but contrary to \zetamp{17}.  Their model had an initial $m=2$ perturbation and an initial faster rotation.  Thus, their initial conditions promoted disc formation and hence they found results similar to our \zetamn{17}, in which the Hall effect is responsible for promoting disc formation.  

\subsection{Counter-rotating envelopes}
During the first core phase, a counter-rotating envelope forms in \zetamn{17}, and at its most massive contains \sm$10^{-3}$~M$_\odot$ and extends to $r \sim 30$~au.  The counter-rotating envelope dissipates with time and disappears completely by \rhoxtwoapprox{2}{-9}, just before the disc forms.  This envelope is smaller and less vertically extended than those found by \citet{Tsukamoto+2015hall,Tsukamoto+2017}.  Their initially stronger magnetic field strength and faster rotation likely required the larger 
envelope at larger radii to conserve angular momentum.  Thus, counter-rotating envelopes are likely a transient feature, with their properties dependent on the Hall effect, ionization rates \citep[e.g.][]{WursterBatePrice2018ion}, and initial conditions.

\section{Summary and conclusion}
\label{sec:conc}

In this study, we followed the collapse of a molecular cloud core through to the formation of the stellar core in a magnetized medium.  We used a self-consistent treatment of non-ideal MHD, and used the canonical cosmic ray ionization rate of \zetaeq{-17}.  We presented models with the magnetic field aligned and anti-aligned to the rotation axis since the Hall effect depends on the magnetic field orientation.   We compared these models to partially ionized models with higher ionization rates (i.e. \zetaeq{-16}), an ideal MHD model and a purely hydrodynamical model.  Our primary conclusions are as follows:
\begin{enumerate}
\item The magnetic braking catastrophe can be solved by the Hall effect if the magnetic field and rotation axis are anti-aligned. During the first core phase, the anti-aligned model with \zetaeq{-17} led to the formation of a gravitationally unstable \sm25~au disc.  The aligned model formed no disc during this phase.  Increasing the cosmic ray ionization rate by a factor of ten yielded models without discs in the first core phase for both magnetic field orientations. 
\item After the second collapse to form a stellar core, the aligned model with \zetaeq{-17} and both models with \zetaeq{-16} formed rotationally supported $1-3$~au discs. No such discs were formed when using ideal MHD.
\item The model with \zetaeq{-17} where the initial magnetic field and rotation vectors are anti-aligned launched a weak $\lesssim 0.3$~\kms first core outflow, while its aligned counterpart launched the fastest (\appx3~\kms) first core outflow amongst our six models. 
\end{enumerate}
By including the Hall effect in non-ideal MHD models that use the canonical cosmic ray ionization rate of \zetaeq{-17}, drastically different results can be produced depending on the initial orientation of the magnetic field.  The Hall effect can qualitatively change the outcome, such that protostars produced from magnetized clouds can resemble results from purely hydrodynamical models (if the initial magnetic field and rotation vectors are anti-aligned) or ideal MHD models (if the vectors are initially aligned).   These results are in agreement with \citet{Tsukamoto+2015hall} who used different initial conditions than presented here, suggesting that our findings are robust and independent of initial conditions, as long as \zetaeq{-17} is used.  Thus we have demonstrated that formation of gravitationally unstable discs with radii more than 25~au is possible despite the presence of magnetic fields. This implies that such discs should indeed exist in the Class 0 phase.

\section*{Acknowledgements}

We would like to thank the referee for useful comments that improved the quality of this manuscript.
JW and MRB acknowledge support from the European Research Council under the European Community's Seventh Framework Programme (FP7/2007- 2013 grant agreement no. 339248).  DJP received funding via Australian Research Council grants FT130100034, DP130102078 and DP180104235.  The calculations for this paper were performed on the DiRAC Complexity machine, jointly funded by STFC and the Large Facilities Capital Fund of BIS  (STFC grants ST/K000373/1, ST/K0003259/1 and ST/M006948/1), and the University of Exeter Supercomputer, a DiRAC Facility jointly funded by STFC, the Large Facilities Capital Fund of BIS, and the University of Exeter.
The research data supporting this publication will be openly available from the University of Exeter's institutional repository.
Several figures were made using \textsc{splash} \citep{Price2007}.  

\bibliography{seventeen.bib}

\label{lastpage}
\end{document}